\documentclass[conference]{IEEEtran}
\IEEEoverridecommandlockouts

\usepackage{cite}

\usepackage[normalem]{ulem} 
\usepackage[colorlinks=true, linkcolor=black, urlcolor=black]{hyperref} 
\usepackage{hyperref}
\usepackage{amsmath,amssymb,amsfonts}
\usepackage{algorithmic}
\usepackage{graphicx}
\usepackage{textcomp}
\usepackage{xcolor}
\usepackage{subcaption}
\usepackage{enumitem}
\usepackage{colortbl} 
\usepackage{array} 
\usepackage{adjustbox}
\def\BibTeX{{\rm B\kern-.05em{\sc i\kern-.025em b}\kern-.08em
    T\kern-.1667em\lower.7ex\hbox{E}\kern-.125emX}}
\begin{document}

\title{MedFocusCLIP : Improving few shot classification in medical datasets using pixel wise attention\\
}

\author{\IEEEauthorblockN{Aadya Arora}
\IEEEauthorblockA{\textit{Department of Electrical Engineering} \\
\textit{Indian Institute Of Technology Gandhinagar}\\
Gujarat, India \\
aadya.arora@iitgn.ac.in}
\and
\IEEEauthorblockN{Vinay Namboodiri}
\IEEEauthorblockA{\textit{Department Of Computer Science} \\
\textit{University Of Bath}\\
Bath, United Kingdom \\
vpn22@bath.ac.uk}
}

\maketitle

\begin{abstract}
With the popularity of foundational models, parameter efficient fine tuning has become the defacto approach to leverage pretrained models to perform downstream tasks. Taking inspiration from recent advances in large language models, Visual Prompt Tuning, and similar techniques, learn an additional prompt to efficiently finetune a pretrained vision foundational model. However, we observe that such prompting is insufficient for fine-grained visual classification tasks such as medical image classification, where there is large inter-class variance, and small intra-class variance. Hence, in this paper we propose to leverage advanced segmentation capabilities of Segment Anything Model 2 \cite{sam2segmentimages} (SAM2) as a visual prompting cue to help visual encoder in the CLIP \cite{clip} (Contrastive Language-Image Pretraining) by guiding the attention in CLIP visual encoder to relevant regions in the image. This helps the model to focus on highly discriminative regions, without getting distracted from visually similar background features, an essential requirement in a fewshot, finegrained classification setting. We evaluate our method on diverse medical datasets including X-rays, CT scans, and MRI images, and report an accuracy of (71\%, 81\%, 86\%, 58\%) from the proposed approach on (COVID, lung-disease, brain-tumor, breast-cancer) datasets against (66\%, 70\%, 68\%, 29\%) from a pretrained CLIP model after fewshot training. The proposed approach also allows to obtain interpretable explanation for the classification performance through the localization obtained using segmentation. 
For demonstrations and visualizations, please visit  \uline{\href{ https://aadya-arora.github.io/MedFocusClip/}{ https://aadya-arora.github.io/MedFocusClip/}}

\end{abstract}

\begin{figure}[t]
    \centering
    \includegraphics[width=0.47\textwidth]{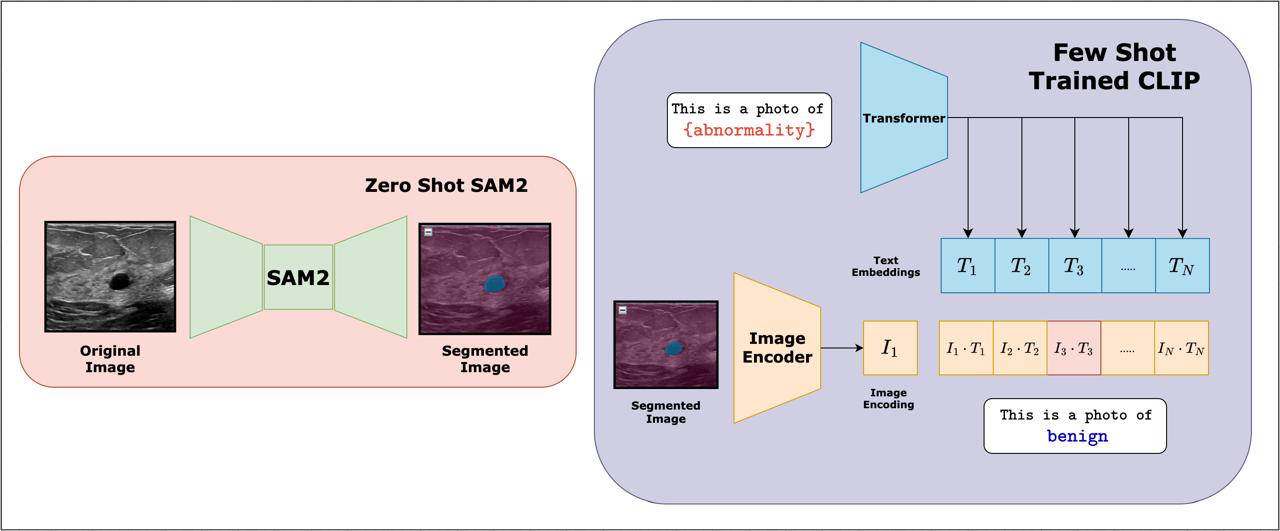} 
    \caption{This image illustrates the proposed MedFocusCLIP framework, combining zero-shot SAM2 segmentation with few-shot trained CLIP for medical image analysis. SAM2 first segments the original image, which is then processed by CLIP's image encoder utilizing a Vision Transformer (ViT) backbone. This ViT-based encoder extracts rich visual features that are aligned with text embeddings, enabling efficient learning and classification from limited medical imaging data. The ViT architecture in the image encoder allows for better handling of global context in medical images, potentially improving the model's ability to detect subtle abnormalities.}
    \label{fig:architecture}
\end{figure}

\section{Introduction}
Automated classification in medical imaging is increasingly necessary to aid healthcare in resource constrained environments without access to specialists. We can obtain such classification using modern deep learning techniques. However, these require access to large annotated datasets in order to work effectively. An alternative approach is to use large foundational models such as CLIP (Contrastive Language Image Pretraining) \cite{clip}. CLIP model has shown impressive zero-shot and few-shot abilities across different domains, including natural imagery and text. However, the challenge remains in the classification performance for specialized domains such as medical imaging, where the inter-class variations are low, and intra-class variance is high. Moreover, such classification through similarity between global visual features and text embedding, lacks interpretability. In this paper, we aim to address these specific challenges.

The focus of this work is to improve the accuracy of few-shot classification for CLIP and provide better interpretability for medical imaging domain. To do so, we consider the substantial advancements made in unsupervised segmentation through recent models like Segment Anything Model 2 (SAM2) \cite{sam2segmentimages}. These models have shown strong ability to localize objects and coherent regions within images. We believe focusing on relevant regions of interest (ROIs), which indicate key anatomical structures or abnormalities, are crucial for accurate classification in medical images. Hence, we propose to leverage SAM2 segmentation mask for prompting CLIP.

We present MedFocusCLIP, a novel framework that combines the advantages of CLIP and SAM2 to enhance few-shot learning for medical image classification. Our method employs SAM2 to create visual prompts that direct CLIP visual encoder's attention to the most pertinent regions of an image, guided by related textual descriptions. This approach allows for a more efficient use of limited labeled data by more precisely aligning relevant visual features with medical concepts. By concentrating on relevant image areas, MedFocusCLIP improves the model's capability to distinguish between subtle features crucial for medical diagnosis.
The key contributions of our work are as follows:
\begin{enumerate}[leftmargin=*]
    \item We introduce a novel architecture that merges visual prompting using SAM2 with the multimodal capabilities of CLIP to enhance medical image classification. 
    \item Our technique refines CLIP’s focus on regions of interest (ROI) in medical images. This leads to better attention mechanisms and more precise classification results. This is particularly advantageous in medical imaging, where accurate localization of abnormalities can significantly influence diagnostic accuracy and patient outcomes.
    \item We validate the effectiveness of our method through extensive experiments on diverse medical datasets including X-rays, CT scans, and MRI images. We report an accuracy of (71\%, 81\%, 86\%, 58\%) on (COVID, lung-disease, brain-tumor, breast-cancer) datasets using our method, against (66\%, 70\%, 68\%, 29\%) from a pretrained CLIP model after fewshot training.
\end{enumerate}

Our research solves the crucial need for data-efficient learning in medical imaging, offering a framework for this problem. Our framework not only pushes the boundaries of few-shot medical image classification but also provides valuable insights into the integration between segmentation models, visual-language models, and domain-specific uses.

\section{Related Works}

\subsection{SAM in Medical Image Segmentation}
SAM has been investigated for its applications in medical image segmentation owing to its advanced capabilities. Studies have concentrated on customizing SAM with fine-tuning methodologies. For example, MedSAM~\cite{Medsam} adjusts the SAM2 mask decoder using extensive medical datasets, whereas SAMed~\cite{SAMed} employs a low-rank adaptation (LoRA)~\cite{hu2021loralowrankadaptationlarge} strategy to train a universal prompt across the dataset images. The SAM Adapter (MSA)~\cite{chen2023samfailssegmentanything} incorporates adapter modules for fine-tuning. These methods have shown significant improvements in performance, frequently equalling or exceeding state-of-the-art fully-supervised models. Nevertheless, these SAM2-based approaches still necessitate large amounts of labeled data for supervised fine-tuning and do not fully exploit the potential of prompt engineering in the model.

Another research direction evaluates the few-shot segmentation capabilities of SAM by using its prompting ability to generate specific object segmentations. Several studies~\cite{sam_cite1, sam_cite2, sam_cite3, sam_cite5, sam_cite4} have assessed SAM's performance on various medical image segmentation tasks using a zero-shot transfer approach. However, effective prompt generation usually requires domain expertise or high-quality labeled data. In contrast, our approach does not rely on supervised training or specialized prompt engineering. 

\subsection{CLIP}
CLIP \cite{clip} is a pre-trained Vision-Language Model (VLM) recognized for its exceptional generalization and zero-shot domain adaptation capabilities. Adaptation of CLIP to various domains often involves prompt engineering, where task-specific semantic details are incorporated \cite{clip}. CLIPSeg \cite{CLIPSeg} enhances CLIP by integrating a transformer-based decoder for dense predictions, while MedCLIP \cite{medclip} expands training data by separately fine-tuning medical images and texts at a lower cost. CXR-CLIP \cite{You_2023} optimizes chest X-ray classification by fine-tuning CLIP's image and text encoders with image-text and image-label datasets. These approaches rely on supervised fine-tuning with medical image-text pairs, often requiring domain expertise for data collection. 

SALip \cite{SALIP} integrates SAM and CLIP for zero-shot medical image segmentation. It uses SAM for exhaustive segmentation and CLIP to retrieve regions of interest, enabling effective organ segmentation without domain-specific prompts or fine-tuning. This approach enhances adaptability across diverse medical imaging tasks.

\section{Proposed System}

\subsection{Overview}

In this study, we introduce \textbf{MedFocusCLIP}, an innovative framework for few-shot medical image classification that leverages visual prompting through the Segment Anything Model 2 (SAM2) and the multimodal capabilities of CLIP (Contrastive Language-Image Pre-training). The framework aims to improve classification performance on medical datasets with limited labeled samples by effectively combining visual and textual information.

\subsection{Methodology}

\begin{figure*}[t]
	\centering
	\begin{subfigure}[t]{0.4\textwidth}
		\centering
		\includegraphics[width=\linewidth]{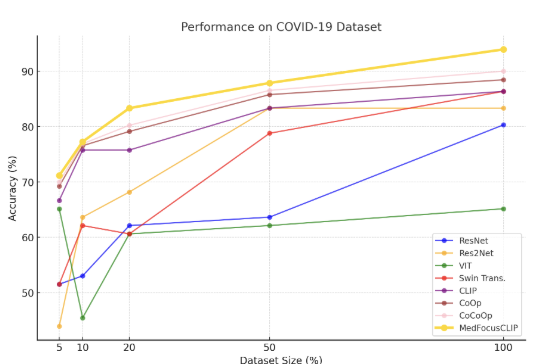} 
		\caption{COVID-19 Dataset}
		\label{fig:covid}
	\end{subfigure}%
	~ 
	\begin{subfigure}[t]{0.4\textwidth}
		\centering
		\includegraphics[width=\linewidth]{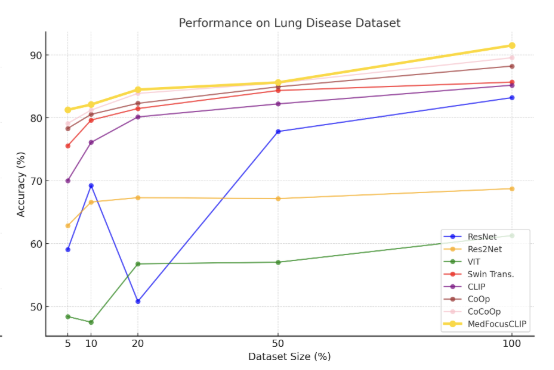} 
		\caption{Lung Disease Dataset}
		\label{fig:lungs}
	\end{subfigure}
	\\ \noindent
	\begin{subfigure}[t]{0.4\textwidth}
		\centering
		\includegraphics[width=\linewidth]{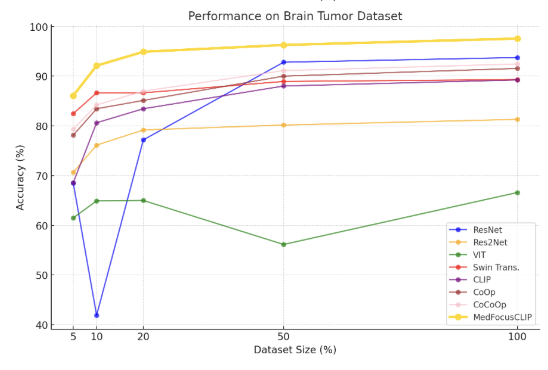} 
		\caption{Brain Tumor Dataset}
		\label{fig:brain}
	\end{subfigure}
	~
	\begin{subfigure}[t]{0.4\textwidth}
		\centering
		\includegraphics[width=\linewidth]{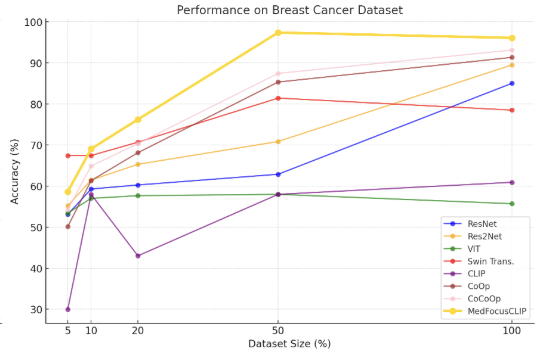} 
		\caption{Breast Cancer Dataset}
		\label{fig:breast}
	\end{subfigure}
	\caption{Performance comparison across different architectures and dataset sizes for different datasets.}
\end{figure*}

\begin{figure}[t]
	\centering
	\begin{subfigure}[t]{0.38\linewidth}
		\centering
		\includegraphics[width=\linewidth,height=1.2\linewidth]{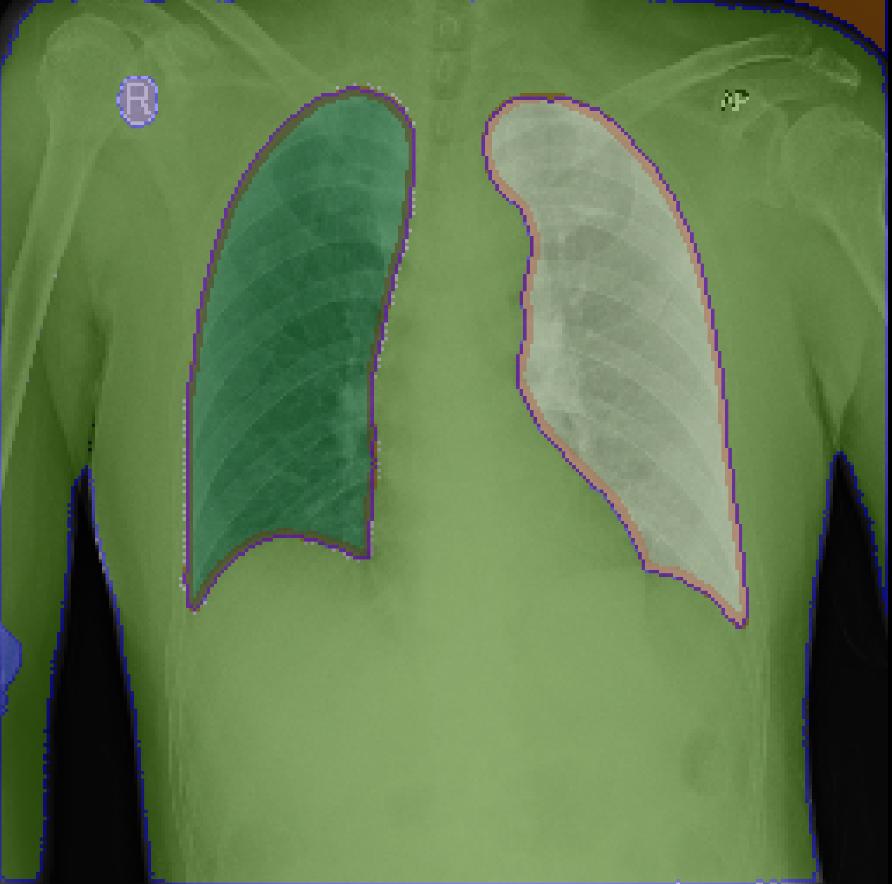} 
		\caption{COVID-19 Dataset}
		\label{fig:covidimg}
	\end{subfigure}%
	~ 
	\begin{subfigure}[t]{0.38\linewidth}
		\centering
		\includegraphics[width=\linewidth,height=1.2\linewidth]{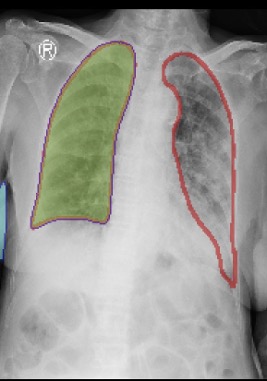} 
		\caption{Lung Disease Dataset}
		\label{fig:lungimg}
	\end{subfigure}
	\\ \noindent
	\begin{subfigure}[t]{0.38\linewidth}
		\centering
		\includegraphics[width=\linewidth,height=\linewidth]{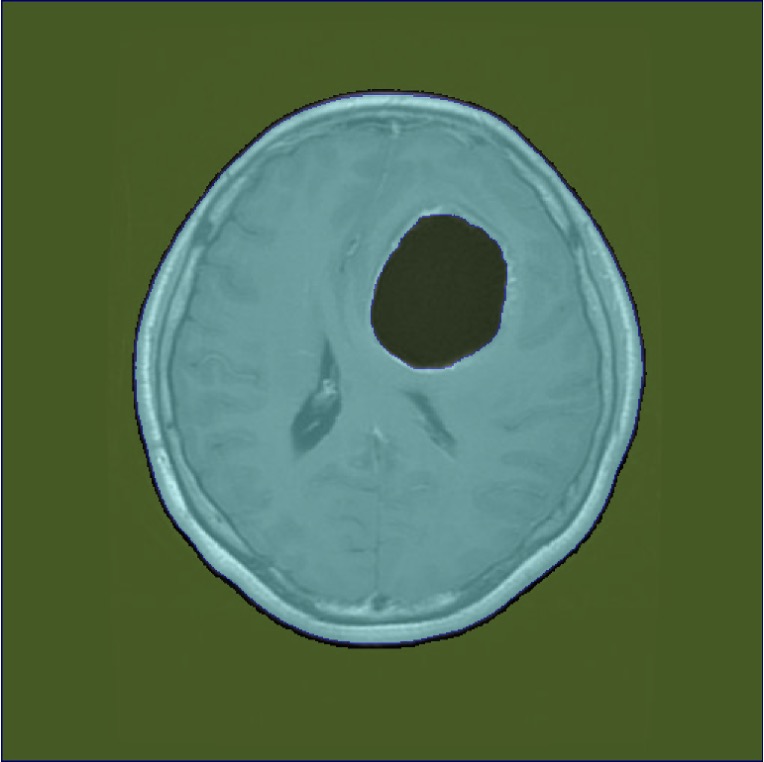} 
		\caption{Brain Tumor Dataset}
		\label{fig:brainimg}
	\end{subfigure}
	~
	\begin{subfigure}[t]{0.38\linewidth}
		\centering
		\includegraphics[width=\linewidth,height=\linewidth]{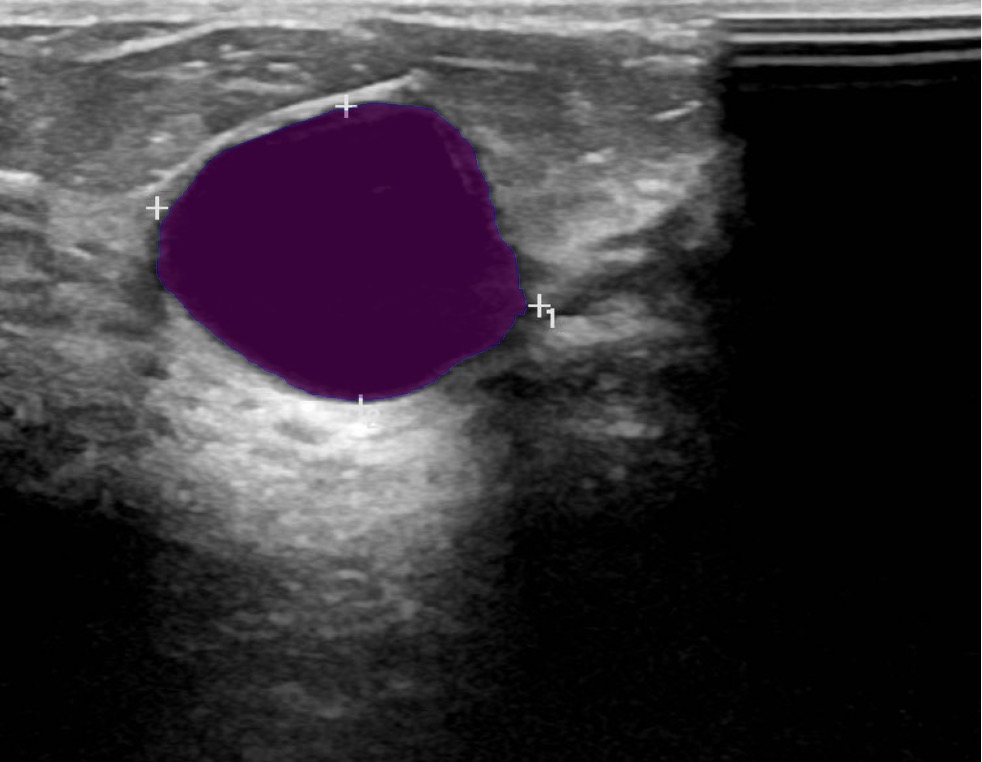} 
		\caption{Breast Cancer Dataset}
		\label{fig:breastimg}
	\end{subfigure}
	\caption{Sample outputs from various datasets showing regions of interest generated from SAM2.}
	\label{fig:sam2output}
\end{figure}

\begin{table}[t]
	\centering
	\small 
	\caption{Performance comparison across different architectures and dataset sizes for \textbf{COVID-19 Dataset}.}
	\resizebox{\linewidth}{!}{
		\begin{tabular}{|>{\columncolor{blue!20}}l|p{1.1cm}|p{1.1cm}|p{1.1cm}|p{1.1cm}|p{1.1cm}|}
			\hline
			\rowcolor{gray!30}
			\textbf{Architecture} & \textbf{5\%} & \textbf{10\%} & \textbf{20\%} & \textbf{50\%} & \textbf{100\%} \\
			\hline
			ResNet \cite{Resnet} & 51.52\% & 53.03\% & 62.12\% & 63.64\% & 80.30\% \\
			Res2Net\cite{Res2Net} & 43.94\% & 63.64\% & 68.18\% & 83.33\% & 83.33\%  \\
			VIT \cite{vit}& 65.15\% & 45.45\% & 60.61\% & 62.12\% & 65.15\% \\
			Swin-Trans \cite{swintransformer} & 51.52\% & 62.12\% & 60.61\% & 78.79\% & 86.36\% \\
			CLIP \cite{clip} & 66.67\% & 75.76\% & 75.76\% & 83.33\% & 86.36\%  \\
			CoOp \cite{zhou2022coop} & 69.23\% & 76.54\% & 79.12\% & 85.78\% & 88.45\% \\
			CoCoOp \cite{zhou2022cocoop} & 70.01\% & 76.89\% & 80.23\% & 86.54\% & 90.01\% \\
			\textcolor{blue}{MedFocusCLIP} & \textcolor{red}{71.15\%} & \textcolor{red}{77.27\%} & \textcolor{red}{83.33\%} & \textcolor{red}{87.88\%} & \textcolor{red}{93.94\%}  \\
			\hline
		\end{tabular}
	}
	\label{tab:COVID-19_Dataset_comparison}
\end{table}

\begin{table}[t]
	\centering
	\caption{Performance comparison across different architectures and dataset sizes for \textbf{Lung Disease Dataset}.}
	\resizebox{\linewidth}{!}{
		\begin{tabular}{|>{\columncolor{blue!20}}l|p{1.1cm}|p{1.1cm}|p{1.1cm}|p{1.1cm}|p{1.1cm}|}
			\hline
			\rowcolor{gray!30}
			\textbf{Architecture} & \textbf{5\%} & \textbf{10\%} & \textbf{20\%} & \textbf{50\%} & \textbf{100\%} \\
			\hline
			ResNet \cite{Resnet} & 59.06\% & 69.23\% & 50.81\% & 77.83\% & 83.21\% \\
			Res2Net \cite{Res2Net} & 62.86\% & 66.62\% & 67.31\% & 67.16\% & 68.74\% \\
			VIT \cite{vit} & 48.40\% & 47.51\% & 56.79\% & 57.04\% & 61.28\% \\
			Swin Trans. \cite{swintransformer}  & 75.56\% & 79.65\% & 81.48\% & 84.35\% & 85.68\% \\
			CLIP \cite{clip}& 70.03\% & 76.10\% & 80.15\% & 82.22\% & 85.19\% \\
			CoOp \cite{zhou2022coop} & 78.32\% & 80.56\% & 82.31\% & 84.94\% & 88.23\% \\
			CoCoOp \cite{zhou2022cocoop} & 79.12\% & 81.24\% & 83.89\% & 85.54\% & 89.56\% \\
			\textcolor{blue}{MedFocusCLIP} & \textcolor{red}{81.28\%} & \textcolor{red}{82.12\%} & \textcolor{red}{84.49\%} & \textcolor{red}{85.63\%} & \textcolor{red}{91.52\%} \\
			\hline
		\end{tabular}
	}
	\label{tab:lung_performance_comparison}
\end{table}

\begin{enumerate}[leftmargin=*]
    \item \textbf{Text Encoding}:
  Given a text prompt \( t \), which can be a class name or a descriptive phrase related to medical conditions, we utilize the text encoder \( E_t \) from CLIP to transform the prompt into a high-dimensional embedding:
    \[
    T = E_t(t) \in \mathbb{R}^d
    \]
    Here, \( T \) represents the encoded text features in a \( d \)-dimensional space, capturing the semantic information relevant to the class label.

    \item \textbf{Image Segmentation and Encoding}:
    To focus on the regions of interest in the medical images, we employ SAM2 to generate segmentation masks. For an input image \( x \), SAM2 outputs a mask \( m \) highlighting the pertinent regions:
    \[
    m = \text{SAM2}(x)
    \]
    We then apply this mask to the image, producing a segmented image \( x_s = x \odot m \), which is passed through CLIP's vision encoder \( E_v \) to obtain image embeddings:
    \[
    I = E_v(x_s) \in \mathbb{R}^{n \times d}
    \]
    The resulting \( I \) captures the visual features of the segmented region. 

    \item \textbf{Multimodal Fusion}:
    We introduce a fusion mechanism \( F \) to effectively combine the text embeddings \( T \) with the image embeddings \( I \). The fusion process is designed to enhance the model's understanding by integrating contextual text information with visual features:
    \[
    M = F(T, I) \in \mathbb{R}^{n \times d}
    \]
    The fused representation \( M \) leverages both modalities. 
    This is crucial for downstream classification tasks.

    \item \textbf{Classification}:
    The fused multimodal embeddings \( M \) are processed by a classification head \( C \), which outputs the probability distribution over possible medical conditions:
    \[
    y = C(M)
    \]
    The classifier is tailored to distinguish between different medical classes based on the combined visual-textual features.

    \item \textbf{Training Objective}:
    The training process involves optimizing a composite loss function that balances two objectives: aligning text and image embeddings through a contrastive loss \( L_c \), and ensuring accurate classification with a cross-entropy loss \( L_e \):
    \[
    L = \lambda L_c + (1 - \lambda) L_e
    \]
    Here, \( \lambda \) is a hyperparameter that controls the trade-off between the two loss components. The contrastive loss encourages the model to align semantically similar text and image pairs, while the cross-entropy loss directly optimizes for classification accuracy.

    \item \textbf{Evaluation}:
    For evaluation, we employ a linear-probe methodology using logistic regression to assess the quality of the learned embeddings. This approach involves training a simple logistic regression classifier on top of the frozen multimodal features to evaluate how well the features separate different classes:
    \[
    y' = \text{LogisticRegression}(M)
    \]

\end{enumerate}

\begin{table}[t]
	\centering
	\caption{Performance comparison across different architectures and dataset sizes for \textbf{Brain tumor Dataset}.}
	\resizebox{\linewidth}{!}{
		\begin{tabular}{|>{\columncolor{blue!20}}l|p{1.1cm}|p{1.1cm}|p{1.1cm}|p{1.1cm}|p{1.1cm}|}
			\hline
			\rowcolor{gray!30}
			\textbf{Architecture} & \textbf{5\%} & \textbf{10\%} & \textbf{20\%} & \textbf{50\%} & \textbf{100\%} \\
			\hline
			ResNet \cite{Resnet} & 68.50\% & 41.88\% & 77.19\% & 92.81\% & 93.75\% \\
			Res2Net \cite{Res2Net} & 70.63\% & 76.13\% & 79.18\% & 80.17\% & 81.31\% \\
			VIT \cite{vit} & 61.48\% & 64.91\% & 64.99\% & 56.14\% & 66.59\% \\
			Swin Transformer \cite{swintransformer} & 82.46\% & 86.65\% & 86.65\% & 88.94\% & 89.32\% \\
			CLIP \cite{clip}& 68.57\% & 80.63\% & 83.45\% & 88.02\% & 89.25\% \\
			CoOp \cite{zhou2022coop} & 78.12\% & 83.45\% & 85.12\% & 90.01\% & 91.56\% \\
			CoCoOp \cite{zhou2022cocoop} & 79.34\% & 84.22\% & 86.98\% & 91.12\% & 92.45\% \\
			\textcolor{blue}{MedFocusCLIP} & \textcolor{red}{86.09\%} & \textcolor{red}{92.12\%} & \textcolor{red}{94.91\%} & \textcolor{red}{96.28\%} & \textcolor{red}{97.57\%} \\
			\hline
		\end{tabular}
	}
	\label{tab:brain_performance_comparison}
\end{table}

\begin{table}[t]
\centering
\caption{ Performance comparison across different architectures and dataset sizes for \textbf{Breast Cancer Dataset}.}
\resizebox{\linewidth}{!}{
\begin{tabular}{|>{\columncolor{blue!20}}l|p{1.1cm}|p{1.1cm}|p{1.1cm}|p{1.1cm}|p{1.1cm}|}
\hline
\rowcolor{gray!30}
\textbf{Architecture} & \textbf{5\%} & \textbf{10\%} & \textbf{20\%} & \textbf{50\%} & \textbf{100\%} \\
\hline
ResNet \cite{Resnet} & 53.09\% & 59.28\% & 60.26\% & 62.87\% & 85.02\% \\
Res2Net  \cite{Res2Net} & 55.20\% & 61.45\% & 65.30\% & 70.85\% & 89.50\% \\
VIT \cite{vit} & 53.42\% & 57.00\% & 57.65\% & 57.98\% & 55.70\% \\
Swin Transformer \cite{swintransformer} & \textcolor{red}{67.43\%} & 67.43\% & 70.68\% & 81.43\% & 78.50\% \\
CLIP  \cite{clip} & 29.97\% & 57.98\% & 43.00\% & 57.98\% & 60.91\% \\
CoOp \cite{zhou2022coop} & 50.12\% & 61.34\% & 68.12\% & 85.34\% & 91.34\% \\
CoCoOp \cite{zhou2022cocoop} & 54.23\% & 64.89\% & 70.34\% & 87.45\% & 93.12\% \\
\textcolor{blue}{MedFocusCLIP}      & 58.63\% & \textcolor{red}{69.06\%} & \textcolor{red}{76.22\%} & \textcolor{red}{97.39\%} & \textcolor{red}{96.09\%} \\
\hline
\end{tabular}
}
\label{tab:breast_cancer_performance}
\end{table}

\section{Results And Discussions}

We evaluate proposed framework in few-shot classification setting using linear probing, and logistic regression. We have used ViT backbone in the CLIP model for all our experiments. 

For COVID19 (Table \ref{tab:COVID-19_Dataset_comparison}) dataset, MedFocusCLIP demonstrated superior performance, achieving an accuracy of 71.15\% with just 5\% of the data and 93.94\% with the full dataset. This highlights its effectiveness in few-shot learning scenarios. CLIP also performed well, particularly with smaller data fractions.
For the lungs dataset (Table \ref{tab:lung_performance_comparison}), MedFocusCLIP again led the performance with 81.28\% accuracy at 5\% data and 91.52\% with the entire dataset. The Swin Transformer showed competitive results but did not surpass MedFocusCLIP in low-data settings.
In the brain dataset (Table \ref{tab:brain_performance_comparison}), MedFocusCLIP achieved 86.09\% accuracy with 5\% of the data and 97.57\% with the full dataset, consistently outperforming other models in few-shot learning.
Lastly, in the breast dataset (Table \ref{tab:breast_cancer_performance}), MedFocusCLIP maintained the highest accuracy, reaching 96.09\% with 100\% of the data and 58.63\% with 5\%.

Figure \ref{fig:sam2output} shows the regions of interest generated by SAM2 for various datasets.

\definecolor{headerColor}{rgb}{0.8, 0.8, 0.8} 
\definecolor{rowColor}{rgb}{0.9, 0.9, 1} 

\subsection{Ablation Study}




\noindent
\subsubsection{Impact of Using SAM Adapter Instead of SAM2}

We replaced SAM2 with the SAM Adapter\cite{chen2023samfailssegmentanything} to evaluate its effectiveness in our framework. The SAM Adapter, designed for better integration with existing models, incorporates task-specific knowledge through visual prompts, which enhances its performance in domains like medical imaging. However, in our tests, the SAM Adapter showed a marginal decrease in segmentation accuracy compared to SAM2. This suggests that while the SAM Adapter is effective in generalizing SAM's capabilities to specialized tasks, SAM2's improved segmentation capabilities helps the CLIP visual encoder to focus better on the relevant regions of interest in medical domain.


\noindent
\subsubsection{Replacing CLIP with SWIN Transformer}

To understand the importance of CLIP encoder in the classification performance, we trained and evaluated our architecture using Swin Transformer, in place of CLIP. The results indicated that although the Swin Transformer achieved satisfactory performance, it did not reach the accuracy level of the original SAM2+CLIP configuration. This highlights the significance of CLIP in handling intricate few-shot classification tasks in medical imaging. The accuracy was 83.41\% which surpasses the 81.43\% accuracy obtained with standard images.

\section{Conclusion}
\label{sec:majhead}

 We introduced MedFocusCLIP, which integrates SAM2 with CLIP for enhanced medical image segmentation. Our experiments, including linear probe evaluations using logistic regression, demonstrated that MedFocusCLIP consistently outperforms other architectures across various medical datasets, particularly in few-shot learning scenarios. The results emphasize MedFocusCLIP's capability to achieve high accuracy even with limited data, making it a crucial tool for medical imaging where annotated data is often scarce. This work highlights the potential of advanced vision-language models like CLIP, paired with task-specific enhancements such as SAM2, to advance the field of medical image analysis.


\bibliographystyle{IEEEbib}
\bibliography{strings,refs}


\end{document}